\documentclass[aps,twocolumn,superscriptaddress,floatfix,longbibliography]{revtex4-2}
\usepackage{amsmath,amssymb,amsthm}
\usepackage{physics}
\usepackage{amsfonts}
\usepackage{mathrsfs}
\usepackage{graphicx}
\usepackage{tabularx}
\usepackage{enumerate}
\usepackage{dcolumn}
\usepackage{bm}
\usepackage[colorlinks,linkcolor=blue,citecolor=blue,urlcolor=blue]{hyperref}

\newcommand{\JD}{J_D}
\newcommand{\QC}{\mathrm{QC}_{\max}}
\newcommand{\QD}{Q_D}
\newcommand{\ii}{\mathrm{i}}
\newcommand{\RR}{\mathrm{RR}}
\newcommand{\RT}{\mathcal{R}\mathcal{T}}

\begin{document}

\title{Quantum phase transitions and quantum-information characterization of a non-Hermitian XY chain with staggered Dzyaloshinskii--Moriya interactions}

\author{Hong Jiang}
\affiliation{Center of Materials Science and Optoelectronics Engineering, College of Materials Science and Opto-Electronic Technology, University of Chinese Academy of Sciences, Beijing 100049, China}

\author{Xiang-Ping Jiang}
\email{2015iopjxp@gmail.com}
\affiliation{School of Physics, Hangzhou Normal University, Hangzhou, Zhejiang 311121, China}

\author{Yan-Chao Li}
\email{ycli@ucas.ac.cn}
\affiliation{Center of Materials Science and Optoelectronics Engineering, College of Materials Science and Opto-Electronic Technology, University of Chinese Academy of Sciences, Beijing 100049, China}

\begin{abstract}

We investigate the phase diagram of a non-Hermitian XY chain with staggered Dzyaloshinskii--Moriya (DM) interactions using quantum-information methods. Through an alternating local-spin-rotation transformation, the model is mapped onto a standard DM-free XY chain, and the resulting transformed Hamiltonian exhibits rotation--time-reversal ($\RT$) symmetry. Combining correlation-function analysis with known phase results of the standard XY chain, we construct the phase diagram for the present system. We further adopt quantum-information-based quantities to systematically assess their performance in characterizing quantum phase transitions within this non-Hermitian system. Our results show that single-site entanglement can only detect the Luttinger-liquid (LL)--paramagnetic (PM) phase transition, which corresponds to the exceptional boundary across which $\RT$ symmetry is restored. In contrast, quantum discord (QD) and quantum coherence (QC) identify both phase boundaries: in addition to locating the exceptional boundary, their second-order derivatives resolve the ferromagnetic (FM)--LL transition inside the $\RT$-broken region. Moreover, measurements of QC along different directions capture the DM-induced relative rotation between the two sublattices, thereby distinguishing the staggered-DM chain from a zero-DM chain with identical effective parameters.

\end{abstract}

\maketitle

\section{Introduction}

Effective non-Hermitian Hamiltonians are widely used to describe open and dissipative systems~\cite{AshidaGongUeda2020,DingFangMa2022,MedenGrunwaldKennes2023}; their eigenvalues can be complex, and their left and right eigenstates generally differ. The spin-$1/2$ $XY$ chain is a widely studied exactly solvable model of quantum magnetism~\cite{LiebSchultzMattis1961,Pfeuty1970,BarouchMcCoyDresden1970,BarouchMcCoy1971}. It also provides a simple setting for studying non-Hermitian quantum systems. When the exchange anisotropy of the $XY$ chain becomes imaginary, the model is non-Hermitian and has rotation--time-reversal ($\RT$) symmetry~\cite{ZhangSong2013}. In the $\RT$-symmetric phase, this antiunitary symmetry keeps the spectrum real, while exceptional points (EPs) mark the coalescence of eigenvalues and eigenvectors~\cite{BenderBoettcher1998,Mostafazadeh2002,Heiss2012,BergholtzBudichKunst2021,WuEtAl2019,NaghilooEtAl2019}. Previous studies showed that the exceptional boundary of this non-Hermitian $XY$ model is a hyperbola and examined the energy, magnetization, and quantum correlations near it~\cite{ZhangSong2013Geometric,MiaoEtAl2024}. In this model, long-distance spin correlations are used to further identify the ferromagnetic (FM), Luttinger-liquid (LL), and paramagnetic (PM) phases: the FM--LL boundary lies within the $\RT$-broken region, whereas the LL--PM boundary coincides with the exceptional boundary and the restoration of $\RT$ symmetry~\cite{LuoMeden2026}. Beyond the imaginary-anisotropy model, other non-Hermitian $XY$ chains, including those with complex or spatially varying fields, have also been studied through their phase diagrams, correlations, fidelity, entanglement, and nonequilibrium dynamics~\cite{LiuXuLi2021,PiLu2021,AgarwalEtAl2024,LiuBatchelor2025,WangEtAl2026}.

The Dzyaloshinskii--Moriya (DM) interaction is an antisymmetric exchange generated by spin--orbit coupling when inversion symmetry is absent at a bond center~\cite{Dzyaloshinsky1958,Moriya1960,YangLiangCui2023}. It favors spin canting and vector chirality~\cite{KatsuraNagaosaBalatsky2005}. Previous studies have demonstrated that uniform DM interactions in non-Hermitian XY chains introduce additional band-tilting terms, which break the symmetry and induce chiral phases~\cite{ZhangEtAl2026}. On the other hand, for staggered DM interactions, the two sublattices of the one-dimensional chain rotate in opposite directions. The DM phase can be absorbed into the effective exchange parameters via spin rotation~\cite{ShekhtmanEntinWohlmanAharony1992}. Staggered DM interactions in Hermitian XY chains have been investigated in earlier works~\cite{MaKong2011,BaranOhanyanVerkholyak2018}. Nevertheless, the effects of staggered DM interactions on quantum states within non-Hermitian XY chains remain largely unexplored.

Quantum-information quantities provide a direct way to examine these phase diagrams~\cite{SuEtAl2024,LiEtAl2024Multipartite}. Entanglement entropy has long been used to study criticality in spin systems and free-fermion chains~\cite{OsterlohEtAl2002,OsborneNielsen2002,VidalEtAl2003,AmicoEtAl2008,PeschelEisler2009}. Quantum discord (QD) describes nonclassical two-site correlations and has been used to locate transitions in Ising, $XXZ$, and Hermitian $XY$ chains with DM interactions~\cite{OllivierZurek2001,HendersonVedral2001,Dillenschneider2008,MazieroEtAl2010,LiuEtAl2011,Sarandy2009}. Quantum coherence (QC) quantifies the asymmetry of a state with respect to a chosen measurement basis and has also been used to locate critical and topological transitions~\cite{BaumgratzCramerPlenio2014,Girolami2014,StreltsovAdessoPlenio2017,KarpatCakmakFanchini2014,LiLin2016}. The maximum quantum coherence $\QC$ has been applied to conventional, Berezinskii--Kosterlitz--Thouless, and topological transitions~\cite{LiZhangLin2019,LvLiLin2022}. Entanglement, fidelity, and dynamical quantities have also been used to study phases, criticality, and exceptional boundaries in non-Hermitian systems~\cite{HerviouRegnaultBardarson2019,TzengEtAl2021,TurkeshiSchiro2023,AgarwalEtAl2024,LuEtAl2024,LiuBatchelor2025,AgarwalEtAl2026}. For the zero-DM non-Hermitian $XY$ chain, entanglement and QD were used to characterize the exceptional boundary~\cite{MiaoEtAl2024}, while QC detected the same boundary and the Hermitian transition at $\gamma=0$~\cite{ZhangYu2025}. However, the phase diagram of a non-Hermitian $XY$ chain with a staggered DM interaction, as well as the capability of quantum-information probes to identify its magnetic and $\RT$ transitions, has not been systematically studied. In particular, whether these probes can detect the magnetic transition inside the $\RT$-broken region remains unclear. Since a local spin rotation can absorb the staggered DM term into effective parameters, a further question arises whether these probes can distinguish the staggered-DM chain from a zero-DM chain with identical effective parameters.

In this work, we determine the phase diagram of a non-Hermitian $XY$ chain with a staggered DM interaction and use quantum-information quantities to probe its magnetic and $\RT$ transitions. We use an alternating local spin rotation to map the model exactly to a zero-DM chain. The staggered interaction is absorbed into the effective exchange, while the chain retains the conjugated $\RT$ symmetry. From this mapping, we obtain the analytic phase boundaries and calculate long-distance spin correlations to identify the FM, LL, and PM phases. We then compare three quantum-information quantities against these boundaries. The single-site entanglement entropy detects only the LL--PM boundary. QD and QC also identify this boundary, and their second derivatives identify the FM--LL transition inside the $\RT$-broken region. They therefore recover the complete FM--LL--PM phase diagram. We also use QC along different directions to reveal the relative rotation between the two sublattices and distinguish the staggered-DM chain from a zero-DM chain with the same effective parameters.

The rest of this paper is organized as follows. Section~\ref{sec:model} presents the model, its exact mapping, and the analytic phase diagram. Section~\ref{sec:probes} discusses how the quantum-information quantities probe the phase boundaries. Section~\ref{sec:conclusion} summarizes the main results.

\section{Model and phase diagram}
\label{sec:model}

\subsection{Exact mapping and magnetic phase diagram}

We consider an even chain with periodic boundary conditions,
\begin{align}
H=-\sum_{j=1}^{N}\Big[&
\frac{J(1+\ii\gamma)}{2}\sigma_j^x\sigma_{j+1}^x
+\frac{J(1-\ii\gamma)}{2}\sigma_j^y\sigma_{j+1}^y
+h\sigma_j^z \nonumber\\
&+(-1)^{j-1}D
(\sigma_j^x\sigma_{j+1}^y-\sigma_j^y\sigma_{j+1}^x)
\Big].
\label{eq:H}
\end{align}
Here $J>0$, $h$ is the transverse field, $\gamma$ is the imaginary anisotropy, and $D$ is the staggered DM coupling. Without loss of generality, we take $h,D\geq0$ in the phase diagrams below. We also take $\gamma\geq0$, since changing the sign of $\gamma$ interchanges the $x$ and $y$ directions without changing the phase boundaries.

Define
\begin{gather}
\phi=\tan^{-1}\!\left(\frac{2D}{J}\right),\qquad
\theta_j=(-1)^{j-1}\frac{\phi}{2},\nonumber\\
U=\prod_j\exp\!\left(-\frac{\ii\theta_j\sigma_j^z}{2}\right).
\label{eq:rotation}
\end{gather}
Under this rotation, the exchange terms depend on the angle difference $\theta_{j+1}-\theta_j$. Our choice of angles cancels the DM phase and changes $J$ to $J_D$. Since neighboring spins rotate in opposite directions, $\theta_j+\theta_{j+1}=0$, and the imaginary-anisotropy term $J\gamma$ remains unchanged. The transformed Hamiltonian is
\begin{align}
H_0=UHU^\dagger=-\sum_j\Big[&
\frac{\JD+\ii J\gamma}{2}\sigma_j^x\sigma_{j+1}^x
\nonumber\\[-2pt]
&+\frac{\JD-\ii J\gamma}{2}\sigma_j^y\sigma_{j+1}^y
+h\sigma_j^z\Big],
\label{eq:Hrot}
\end{align}
with
\begin{equation}
\JD=\sqrt{J^2+4D^2}.
\label{eq:JD}
\end{equation}
For $\gamma>0$, Ref.~\cite{LuoMeden2026} established the FM--LL--PM phase diagram of the zero-DM Hamiltonian $H_0$ from long-distance spin correlations. Earlier studies established the $\RT$ symmetry and exceptional boundary of the same model~\cite{ZhangSong2013,ZhangSong2013Geometric,MiaoEtAl2024}. Using the exact mapping derived in this work, we transfer this known phase structure to the staggered-DM chain and obtain its boundaries in the parameters of Eq.~\eqref{eq:H}. The exchange coupling becomes $\JD$, while the anisotropy amplitude remains $J\gamma$, giving the two phase boundaries
\begin{align}
h_0&=\JD=\sqrt{J^2+4D^2},
\label{eq:h0}\\
h_{\rm EP}&=\sqrt{\JD^2+(J\gamma)^2}.
\label{eq:hEP}
\end{align}

\begin{figure*}[t]
\includegraphics[width=\textwidth]{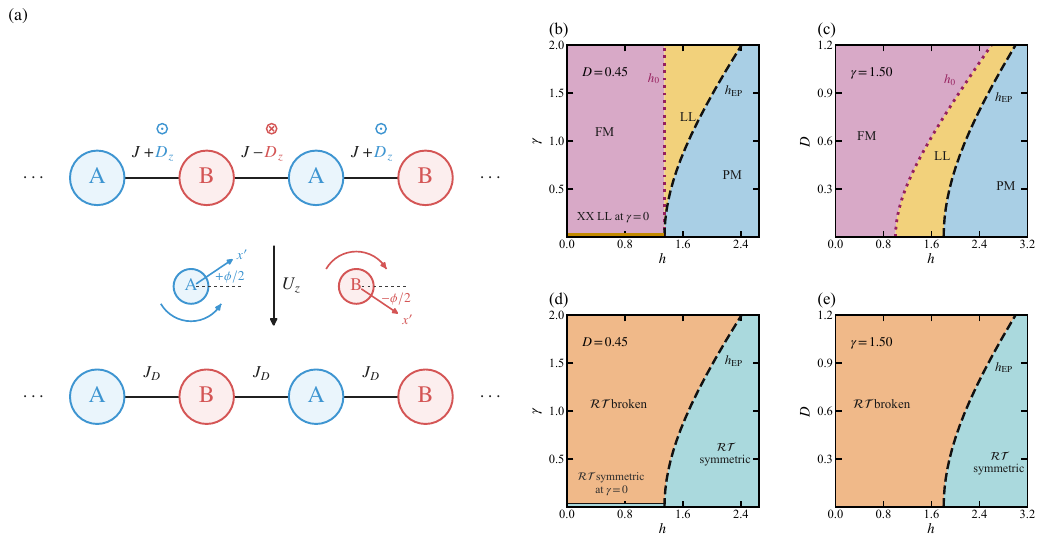}
\caption{Model, exact mapping, and phase diagrams at $J=1$. (a) The alternating rotation removes the $+D,-D$ interaction and changes $J$ to $\JD=\sqrt{J^2+4D^2}$. Panels (b) and (c) show the magnetic phase diagrams for the $(h,\gamma)$ plane at $D=0.45$ and the $(h,D)$ plane at $\gamma=1.50$, respectively. Panels (d) and (e) show the corresponding $\RT$ phase diagrams. At $\gamma=0$, the dark-yellow segment in (b) marks the $XX$ LL phase, while the light-cyan segment in (d) marks the $\RT$-symmetric Hermitian limit. The purple dotted and black dashed curves mark $h_0$ and $h_{\rm EP}$, respectively.}
\label{fig:model_phase}
\end{figure*}

We take the right eigenstate with the smallest $\operatorname{Re}E$ as the ground state and denote it by $|R_0\rangle$. If several eigenvalues have the same minimum real part, we choose the state with the smallest $\operatorname{Im}E$. Expectation values and correlation functions are calculated from the normalized right-right density matrix $\rho_{\RR}=|R_0\rangle\langle R_0|/\langle R_0|R_0\rangle$, as in Ref.~\cite{ZhangYu2025}. The two fields divide the positive-field phase diagram into
\begin{equation}
\begin{aligned}
\mathrm{I:\ FM}:&\quad 0\leq h<h_0,\\
\mathrm{II:\ LL}:&\quad h_0<h<h_{\rm EP},\\
\mathrm{III:\ PM}:&\quad h>h_{\rm EP}.
\end{aligned}
\label{eq:phases}
\end{equation}
Figure~\ref{fig:model_phase}(a) illustrates the alternating local spin rotation that removes the staggered DM interaction. Figures~\ref{fig:model_phase}(b) and~\ref{fig:model_phase}(c) show the magnetic phase diagrams obtained from Eqs.~\eqref{eq:h0} and~\eqref{eq:hEP}. The mapping also explains the $D$ dependence in Fig.~\ref{fig:model_phase}(c). Increasing $|D|$ raises the effective exchange $\JD$, while the anisotropy amplitude $J\gamma$ remains unchanged. Both phase boundaries therefore move to larger $h$. At the same time, $J\gamma/\JD$ decreases, and the LL phase narrows.

To identify these phases directly, we use the exact mapping to calculate the transverse correlation $C_{\RR}^{xx}(r)=\operatorname{Tr}[\rho_{\RR}\sigma_j^x\sigma_{j+r}^x]$ in the rotated frame, following Ref.~\cite{LuoMeden2026}. Figure~\ref{fig:spectrum_correlations}(a) shows that it approaches a nonzero constant in the FM phase and decays as $r^{-1/4}$ at $h_0$. In the LL phase, it has an oscillatory $r^{-3/2}$ envelope. It decays as $r^{-2}$ at $h_{\rm EP}$ and exponentially in the PM phase. The inverse rotation changes the transverse spin axes but preserves the decay, so these results confirm the FM, LL, and PM regions of the staggered-DM chain shown in Figs.~\ref{fig:model_phase}(b) and~\ref{fig:model_phase}(c).

\subsection{Quasiparticle spectrum and the \texorpdfstring{$\RT$}{RT} transition}

The mapped Hamiltonian is invariant under the rotation--time-reversal transformation $\RT$,
\begin{gather}
(\RT)H_0(\RT)^{-1}=H_0,\qquad
{\cal R}=\exp\!\left(-\frac{\ii\pi}{4}\sum_j\sigma_j^z\right),
\nonumber\\[-2pt]
{\cal T}\ii{\cal T}^{-1}=-\ii .
\label{eq:RT_parent}
\end{gather}
Here ${\cal T}$ denotes complex conjugation in the $\sigma^z$ basis. The rotation ${\cal R}$ interchanges the $x$ and $y$ exchange terms after complex conjugation.

Since $H_0=UHU^\dagger$, the staggered-DM Hamiltonian $H$ is invariant under the transformed operator $U^\dagger\RT U$:
\begin{equation}
(U^\dagger\RT U)H(U^\dagger\RT U)^{-1}=H.
\label{eq:RT_original}
\end{equation}
Below we refer to both forms as $\RT$ symmetry~\cite{ZhangSong2013}.

Jordan--Wigner and Fourier transformations reduce $H_0$ to independent two-level blocks,
\begin{equation}
{\cal H}(k)=2\left[(h+\JD\cos k)\tau_z
+\ii J\gamma\sin k\,\tau_x\right],
\label{eq:BdG}
\end{equation}
where $\tau_x$ and $\tau_z$ are Pauli matrices in Nambu space. The quasiparticle energies are
\begin{equation}
\varepsilon(k)=2\sqrt{(h+\JD\cos k)^2-(J\gamma)^2\sin^2 k}.
\label{eq:dispersion}
\end{equation}

Previous work showed that a uniform DM interaction adds a band-tilting term; any nonzero uniform DM coupling breaks the $\RT$ symmetry of the non-Hermitian $XY$ chain and produces chiral phases~\cite{ZhangEtAl2026}. By contrast, the staggered DM interaction studied here is fully absorbed into $\JD$ by the alternating local spin rotation. The spectrum in Eq.~\eqref{eq:dispersion} contains no band-tilting term, and the staggered-DM Hamiltonian retains the conjugated $\RT$ symmetry $U^\dagger\RT U$ in Eq.~\eqref{eq:RT_original}.

The sign of the expression under the square root in Eq.~\eqref{eq:dispersion} determines whether $\varepsilon(k)$ is real or purely imaginary. For $\gamma>0$ and $h<h_{\rm EP}$ [Eq.~\eqref{eq:hEP}], this expression is negative over a finite momentum interval, so $\varepsilon(k)$ is purely imaginary in that interval and $\RT$ symmetry is broken. For $h>h_{\rm EP}$, it is positive for every momentum and the spectrum is entirely real. At $h=h_{\rm EP}$, it is nonnegative for every momentum and vanishes only at $k=\pm k_{\rm EP}$, where
\begin{equation}
k_{\rm EP}=\pi-\tan^{-1}\!\left(\frac{J\gamma}{\JD}\right).
\label{eq:kEP}
\end{equation}
Thus $h=h_{\rm EP}$ is both the LL--PM boundary and the boundary between the $\RT$-broken and $\RT$-symmetric phases, as shown by the LL--PM lines in Figs.~\ref{fig:model_phase}(b) and~\ref{fig:model_phase}(c) and the $\RT$ boundaries in Figs.~\ref{fig:model_phase}(d) and~\ref{fig:model_phase}(e).

At $k=k_{\rm EP}$,
\begin{equation}
{\cal H}(k_{\rm EP})=\frac{2(J\gamma)^2}{h_{\rm EP}}
(\tau_z+\ii\tau_x),\qquad {\cal H}(k_{\rm EP})^2=0.
\label{eq:nilpotent}
\end{equation}
The matrix ${\cal H}(k_{\rm EP})$ is nonzero but its square vanishes. Its two eigenvalues coalesce at zero, leaving only one independent eigenvector. The same holds at $k=-k_{\rm EP}$. These band touchings are therefore exceptional points, and the curve $h=h_{\rm EP}$ defines the exceptional boundary.

At the FM--LL boundary $h=h_0$ [Eq.~\eqref{eq:h0}], a second zero-energy band touching occurs at $k_0=\pi$, where ${\cal H}(k_0)=0$. Unlike the exceptional points at $h_{\rm EP}$, ${\cal H}(k_0)$ has two independent eigenvectors, so the band touching at $h_0$ is diagonalizable. Imaginary quasiparticle energies remain present on both sides of $h_0$. Thus $h_0$ lies inside the $\RT$-broken region.

\begin{figure}[t]
\includegraphics[width=\columnwidth]{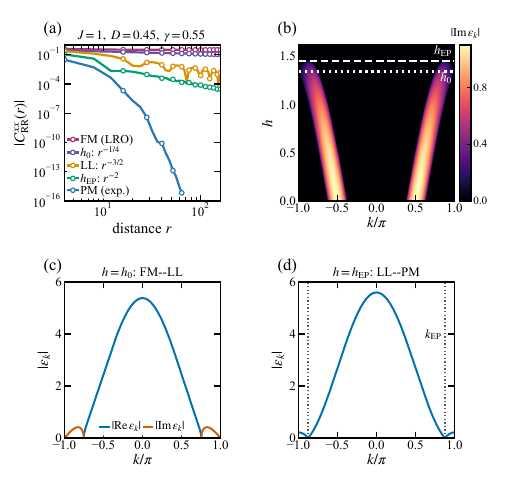}
\caption{Long-distance transverse correlations and quasiparticle spectra under the parameters $J=1$, $D=0.45$, and $\gamma=0.55$. (a) Long-distance transverse correlations calculated in the FM, LL, and PM phases as well as at the two phase boundaries. (b) Color map of $|\operatorname{Im}\varepsilon(k)|$ in the $(k,h)$ plane. The absolute values of the real and imaginary parts of $\varepsilon(k)$, computed at $h_0$ and $h_{\rm EP}$, are presented in panels (c) and (d), respectively.}
\label{fig:spectrum_correlations}
\end{figure}

Figure~\ref{fig:spectrum_correlations}(b) shows that imaginary quasiparticle energies remain present on both sides of $h_0$ and disappear at $h_{\rm EP}$. Figures~\ref{fig:spectrum_correlations}(c) and~\ref{fig:spectrum_correlations}(d) show the diagonalizable band touching at $k_0=\pi$ and the exceptional band touchings at $k=\pm k_{\rm EP}$, respectively.

Together with the long-distance correlations in Fig.~\ref{fig:spectrum_correlations}(a), these spectra show that $h_0$ changes the magnetic correlations while the system remains $\RT$ broken. At $h_{\rm EP}$, the imaginary quasiparticle energies disappear, and the algebraic LL correlations give way to the exponential decay of the PM phase. The FM and LL phases are $\RT$ broken, whereas the PM phase is $\RT$ symmetric.

At $\gamma=0$, the non-Hermitian term in Eq.~\eqref{eq:BdG} vanishes and the spectrum is real for all fields. The entire $\gamma=0$ line is therefore $\RT$ symmetric. The two boundaries merge at $h_0=h_{\rm EP}=\JD$, where the Hermitian $XX$ chain passes from the LL phase to the PM phase without eigenvector coalescence.

\section{Quantum-information probes}
\label{sec:probes}

Based on the phase diagram obtained above, we employ quantum-information quantities to further investigate the magnetic and $\RT$ transitions of the present model. We calculate the single-site entanglement entropy $S_1$, nearest-neighbor quantum discord $\QD$, and QC, and compare their behaviors around $h_0$ and $h_{\rm EP}$. This comparison reveals whether these quantities can identify the magnetic and $\RT$ transitions in this non-Hermitian chain. Of particular interest is the inner FM--LL boundary $h_0$, located within the $\RT$-broken region. A key question addressed in this work is whether these quantum-information quantities can faithfully detect this magnetic transition, given that the spectrum remains complex on both sides of the transition.

All three quantities are calculated from the one- and two-site reduced density matrices of $\rho_{\RR}$. In the thermodynamic limit, these matrices are obtained from the Gaussian fermion correlations of Eq.~\eqref{eq:BdG}. The nearest-neighbor state can be written as
\begin{align}
\rho_{12}=\frac14\Big[&I+
m_z(\sigma_z\otimes I+I\otimes\sigma_z)
+C_{zz}\sigma_z\otimes\sigma_z\nonumber\\
&+\sum_{\mu,\nu=x,y}C_{\mu\nu}
\sigma_\mu\otimes\sigma_\nu\Big],
\label{eq:rho2}
\end{align}
where $m_z=\langle\sigma_j^z\rangle_{\RR}$ and $C_{\mu\nu}=\langle\sigma_j^\mu\sigma_{j+1}^\nu\rangle_{\RR}$. Tracing out either spin gives $\rho_1=(I+m_z\sigma_z)/2$. Thus, $S_1$ depends only on $m_z$, whereas QD and QC also contain the nearest-neighbor spin correlations.

The local rotation $U$ leaves $S_1$ and QD unchanged at the same effective parameters. QC measured along a fixed transverse direction also depends on the rotation of that direction in the laboratory frame. We use this directional dependence below to recover the staggered DM rotation. The two boundaries produce different nonanalytic changes. At $h_{\rm EP}$ the quasiparticle gap closes as $|h-h_{\rm EP}|^{1/2}$, while the leading nonanalytic term at $h_0$ has the form $(h-h_0)^2\ln|h-h_0|$. This difference explains why $h_{\rm EP}$ is already clear in a first derivative and $h_0$ is more easily seen in a second derivative. Appendix~\ref{app:boundaries} gives the corresponding low-energy expansions.

\subsection{Single-site entanglement entropy}

\begin{figure}[t]
  \includegraphics[width=\columnwidth]{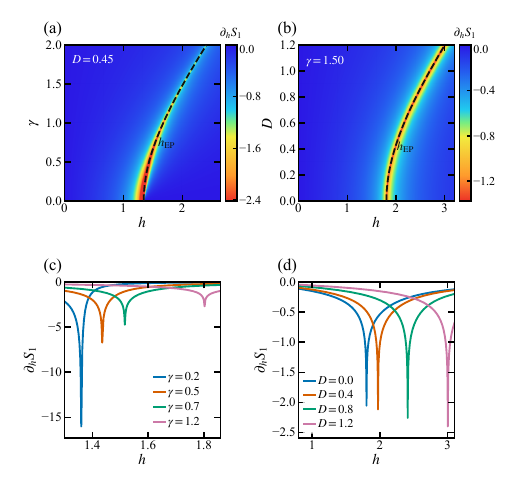}
  \caption{Field derivative of the single-site entanglement entropy at $J=1$. The left column varies $\gamma$ at $D=0.45$, and the right column varies $D$ at $\gamma=1.50$. Heat maps of $\partial_h S_1$ are shown in (a) and (b), with representative field cuts in (c) and (d). The black dashed curves mark $h_{\rm EP}$.}
  \label{fig:entropy}
\end{figure}

For the pure state considered here, the entanglement between one spin and the rest of the chain is given by the von Neumann entropy of the one-site reduced density matrix:
\begin{equation}
S_1=-\operatorname{Tr}(\rho_1\log_2\rho_1)
=H_2\!\left(\frac{1+m_z}{2}\right),
\label{eq:S1}
\end{equation}
where $H_2$ is the binary entropy.
Differentiating Eq.~\eqref{eq:S1} gives
\begin{equation}
\partial_h S_1=\frac12\log_2\!\left(\frac{1-m_z}{1+m_z}\right)\partial_h m_z.
\label{eq:dS1}
\end{equation}

Because $S_1$ depends only on $m_z$, its derivative follows the change of the local $z$ polarization. In both parameter planes, $\partial_h S_1$ develops a sharp minimum along $h_{\rm EP}$, as shown in Figs.~\ref{fig:entropy}(a) and~\ref{fig:entropy}(b). The field cuts in Figs.~\ref{fig:entropy}(c) and~\ref{fig:entropy}(d) show that this minimum follows the analytic $h_{\rm EP}$ as either $\gamma$ or $D$ is varied. The rapid change of $m_z$ as the system enters the PM phase therefore gives a clear local signal of the simultaneous LL--PM and $\RT$ transitions. At $h_0$, the main change occurs in the transverse spin correlations, which do not enter $\rho_1$. Consequently, $S_1$ does not identify the inner FM--LL boundary.

\begin{figure}[t]
  \includegraphics[width=\columnwidth]{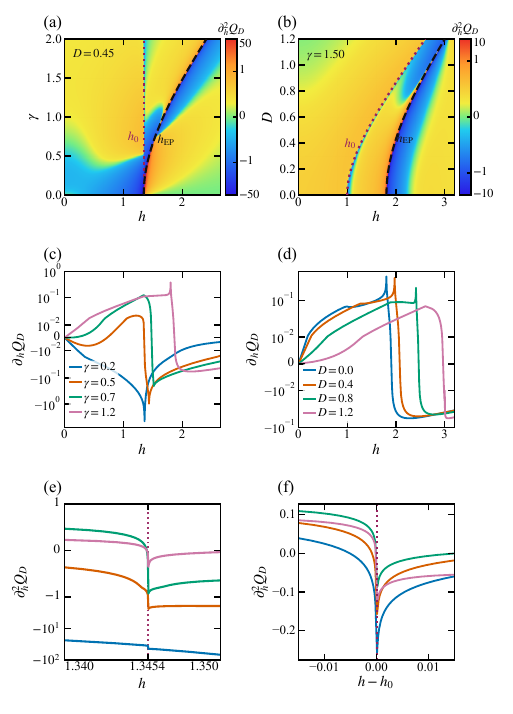}
  \caption{Field derivatives of the nearest-neighbor quantum discord at $J=1$. The left column varies $\gamma$ at $D=0.45$, and the right column varies $D$ at $\gamma=1.50$. Second-derivative maps are shown in (a) and (b), first-derivative cuts in (c) and (d), and second-derivative cuts near $h_0$ in (e) and (f). Purple dotted and black dashed lines mark $h_0$ and $h_{\rm EP}$, respectively.}
  \label{fig:discord}
\end{figure}

\subsection{Quantum discord}

For two neighboring spins, the mutual information is
\begin{equation}
I(\rho_{12})=S(\rho_1)+S(\rho_2)-S(\rho_{12}),
\label{eq:mutual_information}
\end{equation}
where $S(\rho)=-\operatorname{Tr}(\rho\log_2\rho)$ is the von Neumann entropy.
After a projective measurement $\{\Pi_a\}$ on the first spin, the classical correlation is
\begin{equation}
C(\rho_{12})=\max_{\{\Pi_a\}}\left[
S(\rho_2)-\sum_a p_a S(\rho_{2|a})\right].
\label{eq:classical_correlation}
\end{equation}
Here $p_a$ is the probability of outcome $a$, and $\rho_{2|a}$ is the corresponding conditional state of the second spin. The quantum discord is~\cite{OllivierZurek2001,HendersonVedral2001}
\begin{equation}
\QD=I-C.
\label{eq:QD}
\end{equation}

The projected plots for the second-order derivative of the quantum discord (QD) with respect to $h$ at $D=0.45$ and $\gamma=1.50$ are displayed in Figs.~\ref{fig:discord}(a) and~\ref{fig:discord}(b), respectively. The nonanalytic features observed in these plots clearly identify the FM--LL phase transition line $h_0$ (purple dotted line) and the LL--PM phase transition line $h_{\rm EP}$ (black dashed line). As depicted in Figs.~\ref{fig:discord}(c) and~\ref{fig:discord}(d), the first-order derivative of QD with respect to $h$ possesses nonanalytic points that exactly correspond to the $h_{\rm EP}$ phase transition for varying parameters. Nevertheless, no obvious nonanalytic signature occurs at the $h_0$ transition, indicating that the first-order QD derivative fails to capture the FM--LL phase transition.

Further investigations demonstrate that despite the absence of characteristic transition peaks at $h_0$, evident turning points arise and induce prominent peaks in the second-order derivative, as shown in Figs.~\ref{fig:discord}(e) and~\ref{fig:discord}(f). This indicates that even if the energy spectra on both sides of the phase transition remain complex-valued, the second-order derivative of QD can precisely locate the phase-transition point under the breaking of $\RT$ symmetry. Its validity for detecting quantum phase transitions also extends to non-Hermitian systems.

This distinctive performance originates from the inherent properties of quantum discord. Specifically, QD is defined based on the complete two-site quantum state and is susceptible to both local polarization and nearest-neighbor spin correlations. As a result, QD can accurately characterize phase transitions driven by either of the two physical mechanisms.

\subsection{Quantum coherence}
\label{sec:qc}
\label{sec:qcmax}

In this section, we further investigate the performance of another representative quantity, namely QC, for probing quantum phase transitions in non-Hermitian systems.
We adopt the observable-based QC proposed in Ref.~\cite{Girolami2014}.
For a unit vector $\bm n$, the QC associated with the local spin component $\bm n\!\cdot\!\bm\sigma\otimes I$ is
{
\begin{equation}
Q(\bm n)=-\frac14\operatorname{Tr}
\left[\rho_{12},\bm n\!\cdot\!\bm\sigma\otimes I\right]^2.
\label{eq:QC}
\end{equation}
}
It quantifies the extent to which the state $\rho_{12}$ does not commute with the chosen local spin observable $\bm{n}\cdot\bm{\sigma}$. To probe the phase diagram, we evaluate QC along the laboratory $y$ direction on a nearest-neighbor bond starting from the A sublattice and write $\mathrm{QC}_y=Q(\hat{\bm y})$. Substituting Eq.~\eqref{eq:rho2} gives
{
\begin{equation}
\mathrm{QC}_y=\frac14\left(
m_z^2+C_{zz}^2+C_{xx}^2+C_{xy}^2\right).
\label{eq:QCy}
\end{equation}
}
Appendix~\ref{app:coherence} provides the corresponding expressions for the other transverse directions employed below. Since this expression involves both local polarization and nearest-neighbor spin correlations, it should capture both the $h_0$ and $h_{\rm EP}$ quantum phase-transition curves, in the same manner as QD.

\begin{figure}[t]
\includegraphics[width=\columnwidth]{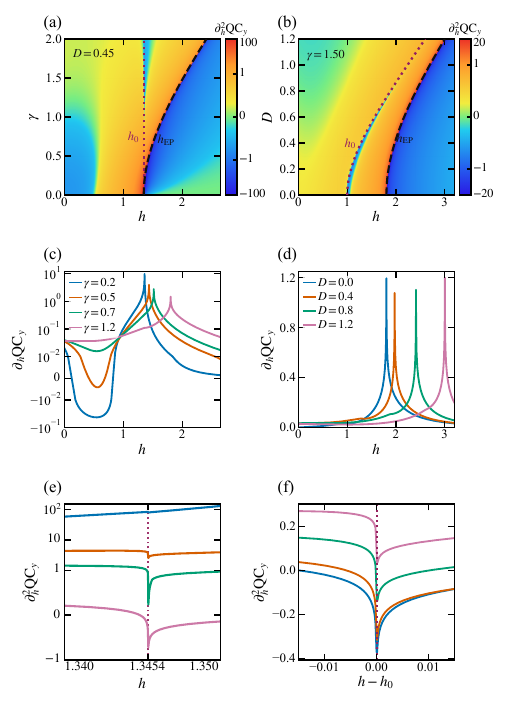}
\caption{Field-derivative quantities of $\mathrm{QC}_y$ at fixed $J=1$. In the left column, $\gamma$ is swept while $D$ is held constant at $0.45$; in the right column, $D$ is swept for fixed $\gamma=1.50$. Panels (a) and (b) show second-derivative maps, (c) and (d) first-derivative cuts, and (e) and (f) second-derivative cuts near $h_0$. The purple dotted and black dashed lines in (a) and (b) indicate $h_0$ and $h_{\mathrm{EP}}$, respectively.}
\label{fig:qc}
\label{fig:qcmax}
\end{figure}

Figures~\ref{fig:qc}(a) and~\ref{fig:qc}(b) display the projected second-order-derivative plots of QC with respect to $h$, computed using the same parameters as Figs.~\ref{fig:discord}(a) and~\ref{fig:discord}(b). Consistent with our earlier conjecture derived from the analysis of Eq.~\eqref{eq:QCy}, the nonanalytic features of QC clearly mark the two phase-transition curves $h_0$ and $h_{\rm EP}$. Similar to QD, the first-order derivative of QC develops singularities at $h_{\rm EP}$ (Figs.~\ref{fig:qc}(c) and~\ref{fig:qc}(d)). By contrast, the FM--LL transition at $h_0$ is unambiguously revealed only through singular peaks in the second-order derivative (Figs.~\ref{fig:qc}(e) and~\ref{fig:qc}(f)). Across different $\gamma$ and $D$, the critical points extracted from these derivative peaks agree with analytical predictions. We thus conclude that QC exhibits efficacy comparable to QD for characterizing the phase diagram of this non-Hermitian system.

\subsection{Quantum coherence and the staggered DM rotation}
\label{sec:directional}

To determine the rotation induced by the staggered DM interaction, we additionally calculate $\mathrm{QC}_x=Q(\hat{\bm x})$ and $\mathrm{QC}_{\pi/4}=Q[(\hat{\bm x}+\hat{\bm y})/\sqrt{2}]$ on bonds beginning on the A and B sublattices. Together with $\mathrm{QC}_y$, these three QC values determine the full angular dependence in the transverse plane:
{
\begin{align}
Q_s(\varphi)={}&\mathrm{QC}_{x}^{(s)}\cos^2\varphi
+\mathrm{QC}_{y}^{(s)}\sin^2\varphi\nonumber\\
&+\left(2\mathrm{QC}_{\pi/4}^{(s)}-\mathrm{QC}_{x}^{(s)}
-\mathrm{QC}_{y}^{(s)}\right)\sin\varphi\cos\varphi .
\label{eq:Qphi_fixed}
\end{align}
}
The transverse coherence angle is therefore
{
\begin{align}
\varphi_s={}&\frac12\operatorname{atan2}\!\Big(
2\mathrm{QC}_{\pi/4}^{(s)}-\mathrm{QC}_{x}^{(s)}-\mathrm{QC}_{y}^{(s)},\nonumber\\[-2pt]
&\mathrm{QC}_{x}^{(s)}-\mathrm{QC}_{y}^{(s)}\Big)\pmod{\pi}.
\label{eq:phi_from_QC}
\end{align}
}
Applying Eq.~\eqref{eq:phi_from_QC} to the two sublattices gives the DM-induced relative rotation.

\begin{figure}[t]
\includegraphics[width=\columnwidth]{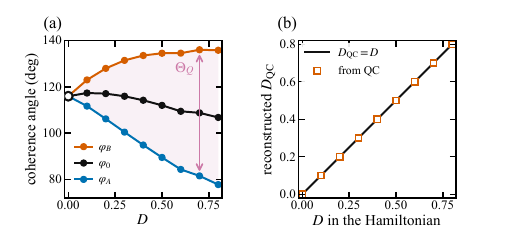}
\caption{Staggered DM rotation reconstructed from QC for $J=1$, $\gamma=0.55$, $h=0.8$, and $N=10$. (a) The angles $\varphi_A$ and $\varphi_B$ on the two sublattices are directly extracted from $\mathrm{QC}_x$, $\mathrm{QC}_y$, and $\mathrm{QC}_{\pi/4}$ through Eq.~\eqref{eq:phi_from_QC}.
The black curve $\varphi_0$ denotes the corresponding angle for the zero-DM chain under identical effective parameters, and $\Theta_Q=\varphi_B-\varphi_A$ characterizes the staggered DM rotation.
(b) Reconstructed DM coupling obtained from $\Theta_Q$ using Eq.~\eqref{eq:ThetaQ}; the black line represents $D_{\mathrm{QC}}=D$.}
\label{fig:directional}
\end{figure}

For the zero-DM chain with the same effective parameters, the two sublattices have the same transverse coherence angle, $\varphi_{A0}=\varphi_{B0}\equiv\varphi_0$. The inverse local rotation shifts this angle in opposite directions:
\begin{equation}
\varphi_A=\varphi_0-\frac{\phi}{2},\qquad
\varphi_B=\varphi_0+\frac{\phi}{2}\pmod{\pi}.
\label{eq:phiAB}
\end{equation}
The common angle $\varphi_0$ changes with the effective parameters as $D$ increases, but it cancels between the two sublattices. Their relative angle is therefore
\begin{equation}
\begin{aligned}
\Theta_Q&=\varphi_B-\varphi_A=\phi,\\
D&=\frac{J}{2}\tan\Theta_Q
=\frac{\JD}{2}\sin\Theta_Q.
\end{aligned}
\label{eq:ThetaQ}
\end{equation}
This relative angle distinguishes the staggered-DM chain from a zero-DM chain with the same effective parameters. Together with $J$ or the effective exchange $\JD$, it also determines the strength of the DM interaction. Figure~\ref{fig:directional}(a) starts at $D=0$, where $\varphi_A=\varphi_B=\varphi_0$, and shows the two angles moving apart as $D$ increases. Each angle is reconstructed directly from $\mathrm{QC}_{x}$, $\mathrm{QC}_{y}$, and $\mathrm{QC}_{\pi/4}$ through Eq.~\eqref{eq:phi_from_QC}. Their separation gives $\Theta_Q$, and Fig.~\ref{fig:directional}(b) shows that Eq.~\eqref{eq:ThetaQ} recovers the DM coupling.

\section{Conclusion}
\label{sec:conclusion}

We investigate the non-Hermitian $XY$ chain with staggered DM interactions and obtain its magnetic and $\RT$ phase diagrams via an exact alternating spin-rotation transformation. This rotation absorbs the staggered interaction into $\JD=\sqrt{J^2+4D^2}$. This feature distinguishes it from uniform DM coupling, which generates band-tilting terms and breaks $\RT$ symmetry. The spectrum after transformation contains no band-tilting terms, and the Hamiltonian with staggered DM terms preserves the conjugate symmetry $U^\dagger\RT U$. Two phase-transition boundaries are obtained through this mapping: $h_0=\JD$ and $h_{\rm EP}=\sqrt{\JD^2+(J\gamma)^2}$. For $\gamma>0$, $h_0$ corresponds to a diagonalizable zero-energy band-touching point, marking the FM--LL transition. At $h_{\rm EP}$, the quasiparticle bands coalesce at the exceptional point; this boundary therefore corresponds simultaneously to the LL--PM transition and the restoration of $\RT$ symmetry. It follows that the FM and LL phases are $\RT$-broken phases, whereas the PM phase respects $\RT$ symmetry.

We further analyze the performance of several typical quantum-phase-transition probes for detecting the two distinct transitions in this non-Hermitian system. The single-site entanglement entropy can only detect $h_{\rm EP}$. Besides locating this exceptional boundary $h_{\rm EP}$, QD and QC can identify the FM--LL transition point $h_0$ within the $\RT$-broken region through their second-order derivatives, thereby recovering the complete FM--LL--PM phase diagram. We offer a plausible physical explanation for how their derivative singularities are able to capture quantum critical points. This broadens the applicability of quantum-information-based probes beyond exceptional boundaries to magnetic transitions where the energy spectrum remains complex-valued on both sides, allowing both classes of transitions in the non-Hermitian phase diagram to be identified. In addition, QC evaluated along the $x$, $y$, and $\pi/4$ directions further allows us to determine the transverse coherence angles on the two sublattices. These sublattice coherence angles satisfy $\tan\Theta_Q=2D/J$, which enables one to distinguish the staggered-DM chain from a zero-DM chain with identical effective parameters, even though the two chains share exactly the same energy spectrum, $S_1$, and QD.

\section*{Acknowledgments}
We acknowledge financial support from National Natural Science Foundation of China (Grant Nos. 12074376 and 12505017), Beijing National Laboratory for Condensed Matter Physics (2025BNLCMPKF017), Beijing Municipal Natural Science Foundation (Grant No. 1222027), and the robotic AI-Scientist platform of Chinese Academy of Sciences.

\appendix

\section{Alternating spin rotation}
\label{app:rotation}

Write $\sigma_j^\pm=(\sigma_j^x\pm\ii\sigma_j^y)/2$. The exchange and DM parts of one bond contain
\begin{align}
&\left[J+2\ii(-1)^{j-1}D\right]\sigma_j^+\sigma_{j+1}^-
+\left[J-2\ii(-1)^{j-1}D\right]\sigma_j^-\sigma_{j+1}^+,
\label{eq:hopping}
\end{align}
while the imaginary anisotropy gives $\ii J\gamma(\sigma_j^+\sigma_{j+1}^+ +\sigma_j^-\sigma_{j+1}^-)$. Under Eq.~\eqref{eq:rotation},
\begin{equation}
U\sigma_j^\pm U^\dagger=e^{\mp\ii\theta_j}\sigma_j^\pm.
\end{equation}
Consequently,
\begin{equation}
\begin{aligned}
U(\sigma_j^+\sigma_{j+1}^-)U^\dagger
&=e^{-\ii(\theta_j-\theta_{j+1})}\sigma_j^+\sigma_{j+1}^-,\\
U(\sigma_j^+\sigma_{j+1}^+)U^\dagger
&=e^{-\ii(\theta_j+\theta_{j+1})}\sigma_j^+\sigma_{j+1}^+.
\end{aligned}
\label{eq:bond_rotation}
\end{equation}
The products with $+\leftrightarrow-$ acquire the opposite phases, and the field term is unchanged because $U\sigma_j^zU^\dagger=\sigma_j^z$. Writing
\begin{equation}
J+2\ii(-1)^{j-1}D=\JD e^{\ii(-1)^{j-1}\phi},
\end{equation}
and using $\theta_j-\theta_{j+1}=(-1)^{j-1}\phi$, the first phase in Eq.~\eqref{eq:bond_rotation} cancels the DM phase and changes both exchange coefficients to $\JD$. The sum $\theta_j+\theta_{j+1}=0$ leaves the pairing terms unchanged. This gives Eq.~\eqref{eq:Hrot}.

The inherited antiunitary symmetry follows immediately:
\begin{align}
{\cal A}_D H{\cal A}_D^{-1}
&=U^\dagger{\cal A}_0(UHU^\dagger){\cal A}_0^{-1}U\nonumber\\
&=U^\dagger H_0U=H.
\end{align}
It also satisfies ${\cal A}_D^2=1$.

\section{Low-energy forms at the two boundaries}
\label{app:boundaries}

Near the exceptional boundary, set $h=h_{\rm EP}+\delta h$ and $k=k_{\rm EP}+q$. The radicand in Eq.~\eqref{eq:dispersion} becomes
\begin{equation}
(h+\JD\cos k)^2-(J\gamma)^2\sin^2 k
=(J\gamma)^2\left(q^2+\frac{2\delta h}{h_{\rm EP}}\right)+\cdots .
\label{eq:EP_expansion}
\end{equation}
The quasiparticle gap therefore closes with the square-root form $|\delta h|^{1/2}$.

Near the inner boundary, set $h=h_0+\delta h$ and $k=k_0+q$, where $h_0=\JD$ and $k_0=\pi$. Then
\begin{align}
&(h+\JD\cos k)^2-(J\gamma)^2\sin^2 k\nonumber\\
&\quad=\delta h^2+
\left[\JD\delta h-(J\gamma)^2\right]q^2\nonumber\\
&\qquad+\left[\frac{\JD^2}{4}+\frac{(J\gamma)^2}{3}
-\frac{\JD\delta h}{12}\right]q^4+O(q^6).
\label{eq:h0_expansion}
\end{align}
The lower integration edge is $q_{\min}=|\delta h|/(J\gamma)+O(\delta h^2)$. A representative anomalous contraction contains
\begin{align}
{\cal I}(\delta h)\propto\int_{q_{\min}}^\Lambda dq\,
\sqrt{(J\gamma)^2q^2-
\left(\delta h+\frac{\JD}{2}q^2\right)^2},
\end{align}
whose nonanalytic contribution is $\delta h^2\ln|\delta h|$. This produces the singularity in the second derivatives of QD and $\mathrm{QC}_y$ at $h_0$.

\section{Quantum coherence and DM rotation}
\label{app:coherence}

The rotation can be derived directly from QC along three directions. Define
{
\begin{equation}
A_x=[\rho_{12},\sigma_x\otimes I],\qquad
A_y=[\rho_{12},\sigma_y\otimes I].
\label{eq:Axy}
\end{equation}
}
For a transverse direction $\bm n=(\cos\varphi,\sin\varphi,0)$, the commutator in Eq.~\eqref{eq:QC} is $A_x\cos\varphi+A_y\sin\varphi$. Therefore,
{
\begin{align}
Q(\varphi)={}&\mathrm{QC}_x\cos^2\varphi
+\mathrm{QC}_y\sin^2\varphi\nonumber\\
&-\frac12\operatorname{Tr}(A_x A_y)
\sin\varphi\cos\varphi .
\label{eq:Q_fixed_derivation}
\end{align}
}
The QC along the $\pi/4$ direction is
{
\begin{equation}
\mathrm{QC}_{\pi/4}=\frac{\mathrm{QC}_x+\mathrm{QC}_y}{2}
-\frac14\operatorname{Tr}(A_x A_y).
\label{eq:QCpi4_cross}
\end{equation}
}
Eliminating $\operatorname{Tr}(A_x A_y)$ between Eqs.~\eqref{eq:Q_fixed_derivation} and~\eqref{eq:QCpi4_cross} gives Eq.~\eqref{eq:Qphi_fixed}. The larger of its two stationary values occurs at
{
\begin{align}
2\varphi={}&\operatorname{atan2}\!\Bigl(
2\mathrm{QC}_{\pi/4}-\mathrm{QC}_x-\mathrm{QC}_y,\nonumber\\[-2pt]
&\mathrm{QC}_x-\mathrm{QC}_y\Bigr)\pmod{2\pi},
\label{eq:phi_fixed_derivation}
\end{align}
}
which gives Eq.~\eqref{eq:phi_from_QC}. Substituting Eq.~\eqref{eq:rho2} also gives
{
\begin{align}
\mathrm{QC}_x&=\frac14(m_z^2+C_{zz}^2+C_{yx}^2+C_{yy}^2),\nonumber\\
\mathrm{QC}_y&=\frac14(m_z^2+C_{zz}^2+C_{xx}^2+C_{xy}^2),\nonumber\\
2\mathrm{QC}_{\pi/4}-\mathrm{QC}_x-\mathrm{QC}_y
&=-\frac12(C_{xx}C_{yx}+C_{xy}C_{yy}).
\label{eq:fixed_QC_elements}
\end{align}
}

Under the local rotation in Eq.~\eqref{eq:rotation}, the angular dependence on site $j$ satisfies
{
\begin{equation}
Q_{\rm lab}^{(j)}(\varphi)=Q_0(\varphi+\theta_j).
\label{eq:Q_rotation}
\end{equation}
}
Its coherence angle therefore changes by $-\theta_j$. With $\theta_A=\phi/2$ and $\theta_B=-\phi/2$, this gives Eqs.~\eqref{eq:phiAB} and~\eqref{eq:ThetaQ}.

For the calculation in Fig.~\ref{fig:directional}, we diagonalize the Hamiltonian in Eq.~\eqref{eq:H} for $N=6,8,10$, $J=1$, $\gamma=0.55$, and $h=0.8$; the figure shows the $N=10$ data. The two-site density matrices are taken on bonds beginning on the A and B sublattices. We calculate $\mathrm{QC}_{x}$, $\mathrm{QC}_{y}$, and $\mathrm{QC}_{\pi/4}$ on each bond. These three QC values give $\varphi_A$ and $\varphi_B$ through Eq.~\eqref{eq:phi_from_QC} and $D_{\rm QC}$ through Eq.~\eqref{eq:ThetaQ}.

\bibliography{NonHXY-DM-QC}

\end{document}